\documentclass[tighten,twocolumn]{aastex62}

\usepackage{amsmath,amssymb}

\newcommand{\kepler}{\textit{Kepler}}

\received{January 1, 2018}
\revised{January 7, 2018}
\accepted{\today}
\submitjournal{ApJL}

\shorttitle{LISA Circumbinary Exoplanets}
\shortauthors{Steffen, Wu, and Larson}

\begin{document}

\title{Detecting circumbinary exoplanets and hierarchical stellar triples with the LISA gravitational radiation mission}

\correspondingauthor{Jason H. Steffen}
\author{Jason H. Steffen}
\email{jason.steffen@unlv.edu}
\affiliation{Department of Physics and Astronomy\\
University of Nevada, Las Vegas\\
Las Vegas, NV 89154}

\author{Dong-Hong Wu}
\affiliation{Department of Physics and Astronomy\\
University of Nevada, Las Vegas\\
Las Vegas, NV 89154}
\affiliation{School of Astronomy and Space Science\\
Key Laboratory of Modern Astronomy and Astrophysics in Ministry of Education\\
Nanjing University\\
Nanjing 210093, China}

\author{Shane L. Larson}
\affiliation{Center for Interdisciplinary Exploration and Research in Astrophysics\\
Northwestern University\\
Evanston, IL 60208}



\begin{abstract}

We investigate the possibility of detecting planetary or stellar companions orbiting white dwarf binaries using the LISA gravitational radiation detector.  Specifically, we consider the acceleration of the barycenter of the white dwarf binary due to the orbiting third body as well as the effect of changes in the tidal field across the binary due to the perturber's eccentric orbit.  We find that the movement of the barycenter is detectable for both stellar and planetary mass objects.  If circumbinary planets occur with frequencies similar to gas giant planets around isolated main sequence stars, then we expect to find of order 10 such planets in four years of LISA observations.  For a longer, ten-year mission the accessible parameter space for planetary mass, orbital period, and binary orbital period grows and LISA's associated yield increases to $\sim 100$ expected detections.

\end{abstract}



\section{Introduction}

Among the groundbreaking discoveries from NASA's \kepler\ mission was the observation of circumbinary planets---planets orbiting an inner binary star \citep{Doyle:2011}.  To date, there are nearly two dozen such planets identified in nearly as many systems \citep{Welsh:2012,Qian:2012,Orosz:2012b,Orosz:2012a,Welsh:2015,Kostov:2014,Lee:2014,Kostov:2016}.  These discoveries show that such planetary systems exist, but it remains difficult to determine the occurrence rates of circumbinary planets.

A major challenge to establishing the frequency of these planets, especially using transit data, is that the transit signature in a binary system is constantly changing.  Thus, one cannot analyze the lightcurve using standard techniques where the data are folded at a fixed orbital period to cause the transits to overlap.  Rather, single systems are identified (often by eye) and the lightcurve is studied in detail to tease out additional transits in the changing flux from the dynamically evolving stellar system.

The anticipated Laser Interferometer Space Antenna mission (LISA) may provide an additional means to study the circumbinary planet population.  Some work has been done to examine the gravitational radiation (or gravitational wave, GW) signature of planets orbiting a central star \citep{Cunha:2018}.  They find that some of the more massive, and shortest-period exoplanets could be detected via the direct gravitational radiation they emit with their host star.  Here, by contrast, we investigate the possibility of detecting planets orbiting short-period, double white dwarf (DWD) stars.

The galaxy is expected to harbor $\sim 10^{7}$ ultra-compact binaries that radiate gravitational waves in the LISA band \cite{Hils:2000,Nelemans:2001}; the vast majority of these are double white dwarfs.  Of those binaries, many have similar orbital frequencies and their overlapping GW signals will cause an irreducible foreground noise or confusion limit.  From the total population of ultra-compact binaries, data analysis studies have modeled the population of binaries that LISA should be able to individually resolve above the confusion foreground.  The expectation is that there will be a few $\times 10^{4}$ such systems \cite{CrowderCornish:2007} with strong GW signatures due to their relatively slow orbital decay (when compared to binary black holes or neutron stars).

Should planets or stellar companions orbit these DWDs, their gravitational perturbations would cause periodic shifts in the GW signal.  We investigate the possibility of detecting the GW signature of a companion object orbiting a DWD via two different kinds of interactions, the Doppler shift about the system barycenter and orbital period variation from a changing tidal field.

This paper is organized as follows.  In Section \ref{signal} we examine the modulation of a GW signal that is caused by these two different mechanisms.  In Section \ref{population} we estimate the number and distribution of DWDs that would be suitable targets for planet detection and make an estimate for the detection probability given our galactic model for DWDs and the expected GW signature from circumbinary companions.  Finally, we give our concluding remarks in Section \ref{conclude}.

\section{Expected Gravitational Radiation Signal}\label{signal}

We examine two frequency-modulating signals in the GW signature of a binary source caused by an orbiting third body.  The first signal is the acceleration of the GW source about the barycenter by the planet---the variation of the Doppler shift.  The second signal is a direct modulation of the orbital period of the GW source induced by the changing tidal field of a distant companion on an eccentric orbit.  Each of these sources is discussed in turn.

\subsection{Acceleration about the barycenter}

The acceleration of the GW source about the system barycenter produces a Doppler modulation of the GW frequency.  To leading order in $v/c$, where $v$ is the radial velocity of the source, the Doppler effect in frequency (relative to the Barycenter of the solar system) is given by
\begin{equation}
f=f_0\left(1+\frac{v}{c}\right)
\end{equation}
where $f$ is the observed orbital frequency of the GW source and $f_0$ is the orbital frequency of the source in the absence of a perturbing object (twice the orbital frequency of the DWD).  With a planetary object (or stellar companion) orbiting the DWD, the amplitude of the radial velocity signal is given by
\begin{equation}
v=\left(\frac{m}{M_{\star}+m}\right)\sqrt{\frac{G(m+M_{\star})}{a}}\sin i
\end{equation}
where $m_\star$ is the mass of the binary, $m$ is the mass of the perturbing companion, $a$ is the orbital semi-major axis, $G$ is Newton's constant, and $i$ is the inclination of the orbit with respect to the line of sight.

The expected frequency shift amplitude of the DWD is
\begin{equation}
\Delta f \simeq \frac{f_0}{c}\left(\frac{m}{M_{\star}+m}\right)\sqrt{\frac{G(M_\star+m)}{a}}\sin i.
\end{equation}
Or, in terms of the planetary orbital period $P$,
\begin{equation}
\Delta f \simeq f_0 \left(\frac{m}{M_\star +m}\right) \left( \frac{2\pi G (M_\star+m)}{c^3 P} \right)^{1/3} \sin i.
\end{equation}
The average value of $\sin i$ is $\pi/4$ assuming that the orbital angular momentum of the planets are uniformly distributed in space, so the average expected amplitude of the signal will be:
\begin{equation}
\begin{split}
\Delta f &= 7.42\times 10^{-8}\text{Hz} \\ &
\times \biggl( \frac{f_0}{\text{Hz}} \biggr) \biggl( \frac{m}{m_{\text{J}}} \biggr)\biggl( \frac{M_\star+m}{M_\odot} \biggr)^{-2/3}\biggl( \frac{P}{\text{year}} \biggr)^{-1/3}.\label{dp}
\end{split}
\end{equation}
For a four-year LISA mission, with a frequency resolution of $\sim 7.92\times 10^{-9}$ Hz, the minimum detectable change of the orbital frequency is $3.96\times10^{-9}$ Hz (since $f_{\rm GW}=2f$ \citep{Postnov2014}).

\begin{figure}
\includegraphics[width=0.45\textwidth]{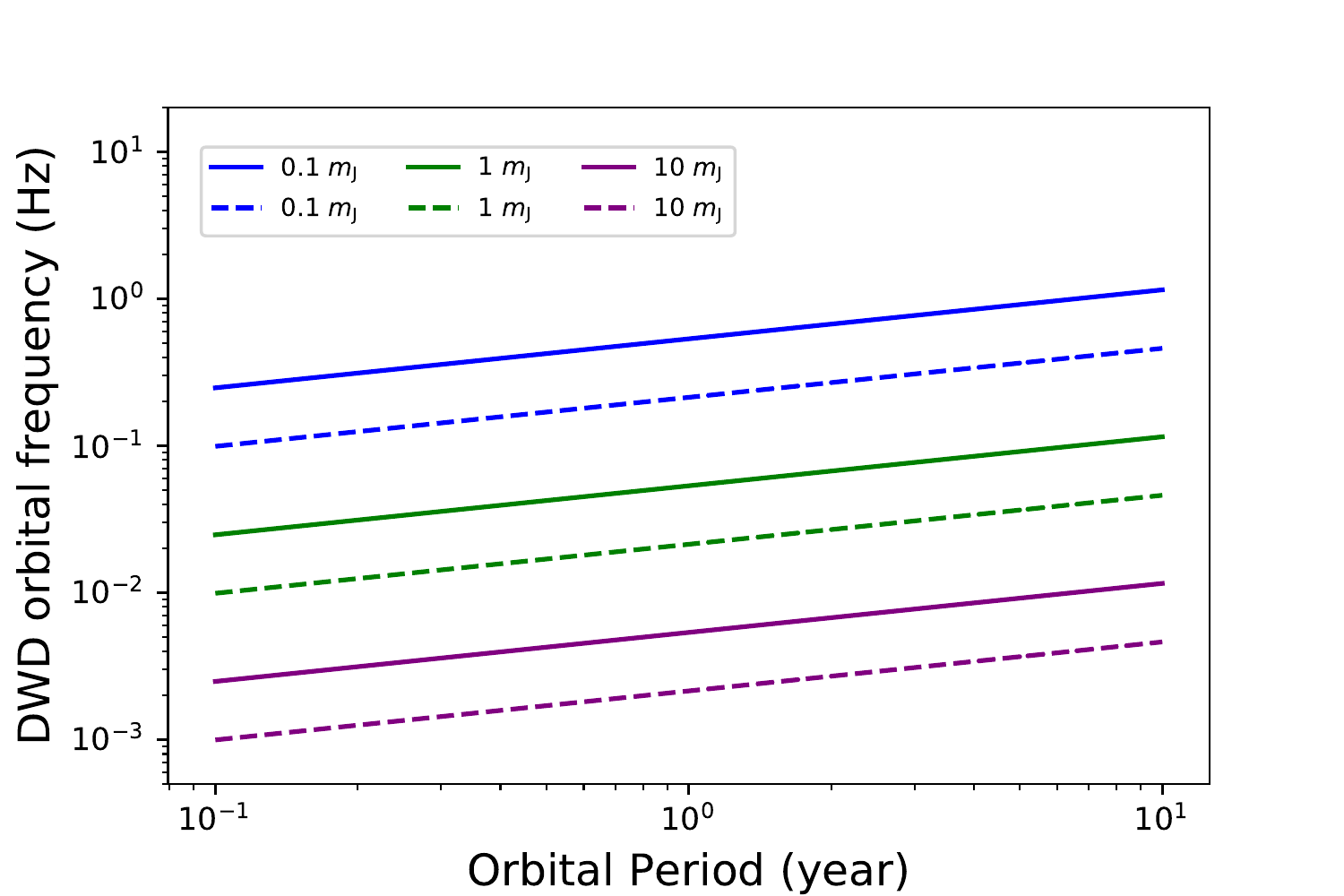}
\caption{DWD orbital frequency where a planet can be detected in a four-year (solid lines) and ten-year (dashed lines) LISA mission (i.e. a frequency resolution of $7.92\times10^{-9}$ Hz and $3.16\times 10^{-9}$ Hz) as a function of planet orbital period for planets with masses of 0.1 $M_{\text{J}}$ (blue), 1.0 $M_{\text{J}}$ (green), and 10 $M_{\text{J}}$ (purple). This plot assumes a total DWD mass of 1 $M_\odot$. \label{freqvsper}}
\end{figure}

The orbital frequency of the DWD needed to detect a planet with a given orbital period is shown in Figure \ref{freqvsper}.  We see that gas giant planets on year-long orbits can produce a detectable signal if the DWDs have orbital frequencies of order 10 milliHertz.  DWDs with these orbital frequencies are expected throughout the galaxy according to galactic population models \citep{Nelemans:2001}.  We will examine the number of possible detections we can expect using this mechanism below for both a 4-year and a 10-year LISA mission below.

\subsection{Tidal modulation of orbital periods}

If the third body is on a wide eccentric orbit, the changing proximity to the DWD will modulate the binary's orbital period.  When the perturbing object is near the binary, the binary orbital period increases.  Then, as the perturbing object retreats to its apocenter, the binary orbital period decreases---approaching its unperturbed value depending upon the apocentric distance.  This scenario was derived for triple star systems in \citet{Borkovits:2003} and \citet{Borkovits:2011}.  To simplify the problem, we assume the orbit of the inner binary is circular and the perturbing object is coplanar with the binary (as was done in \citet{Agol:2005}.  If we consider the time between eclipses of the binary, the deviations from a constant orbital period (O-C) of the eclipsing time $t$ is given by:
\begin{equation}
\begin{split}
O-C &= \frac{1}{2\pi \left(1-e^2_3\right)^{3/2}} \frac{P^2_{12}}{P_3} \frac{m_3}{m_1+m_2+m_3} \\ & \times \left( f_3 -M_3 + e_3\sin f_3 \right)
\end{split}
\end{equation}
where $P_{12}$ is the orbital period of the DWD, and $e_3$, $f_3$, and $M_3$ are the eccentricity, true anomaly, and mean anomaly of the perturbing object, respectively.  The orbital elements are referred to the plane perpendicular to the line of sight and going across the center of mass of the binary.

The departure of the transiting period $t_{j+1}-t_{j}$ from a constant period,  where $t_{j}$ is the $j$th eclipsing time, is therefore:
\begin{widetext}
\begin{equation}
\Delta P =\frac{1}{\left(1-e_3^2\right)^{3/2}}\frac{P_{12}^3}{P^2_3}\frac{m_3}{m_1+m_2+m_3} \biggl(3e_3\left(1-\frac{3}{8}e^2_3\right)\cos M_3 + \frac{9}{2}e^2_3\cos 2M_3+\frac{53}{8}e^3_3\cos 3M_3+O\left(e^4_3\right)\biggr)
\end{equation}
In frequency space this equation is approximately:
\begin{equation}
\Delta f =\frac{1}{(1-e_3^2)^{3/2}}\frac{P_{12}}{P^2_3}\frac{m_3}{m_1+m_2+m_3}\biggr(3e_3\left(1-\frac{3}{8}e^2_3\right)\cos M_3+
\frac{9}{2}e^2_3\cos 2M_3+\frac{53}{8}e^3_3\cos 3M_3+O\left(e^4_3\right)\biggr) 
\label{eq1}
\end{equation}
\end{widetext}

\begin{figure}
\includegraphics[width=0.5\textwidth]{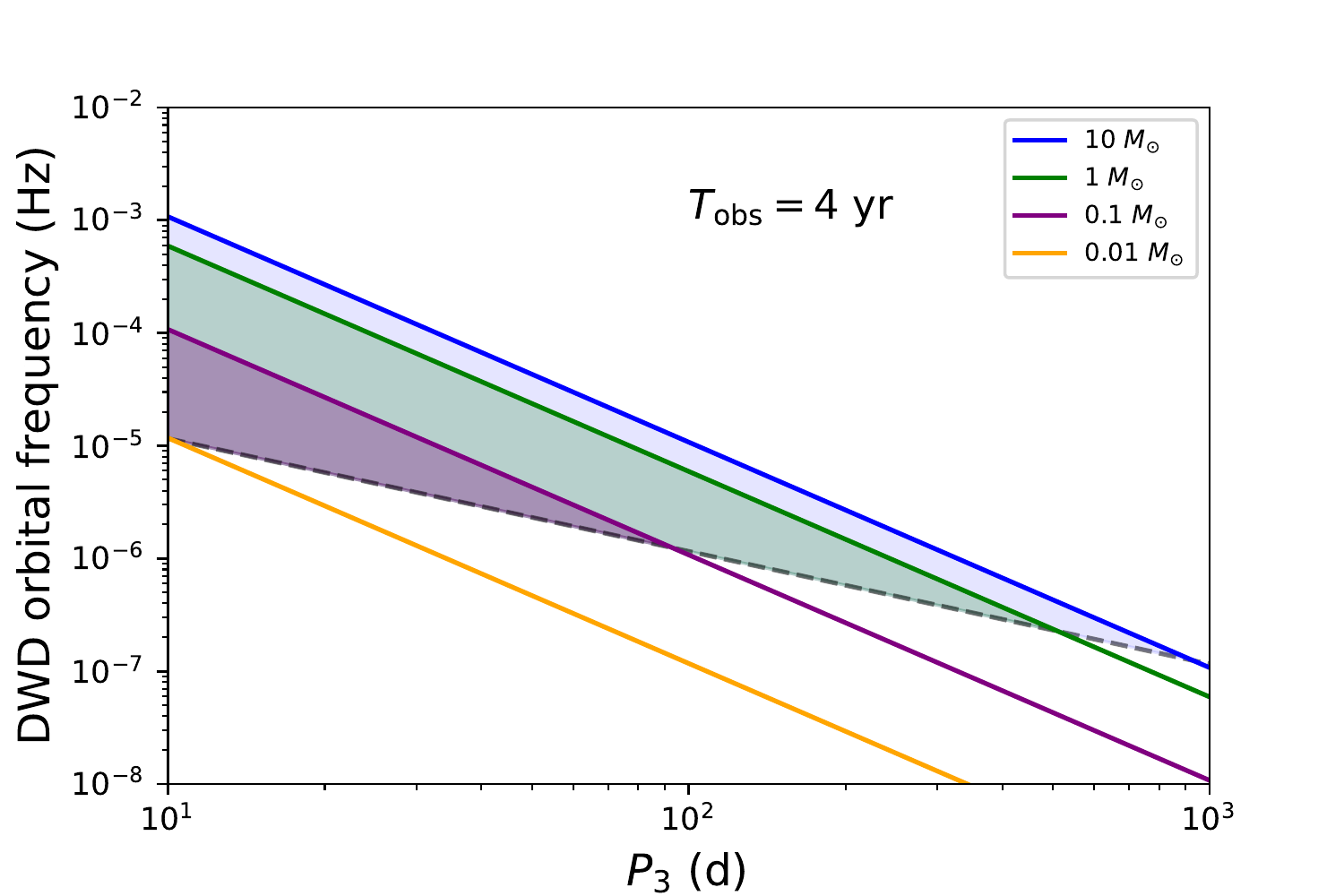}
\includegraphics[width=0.5\textwidth]{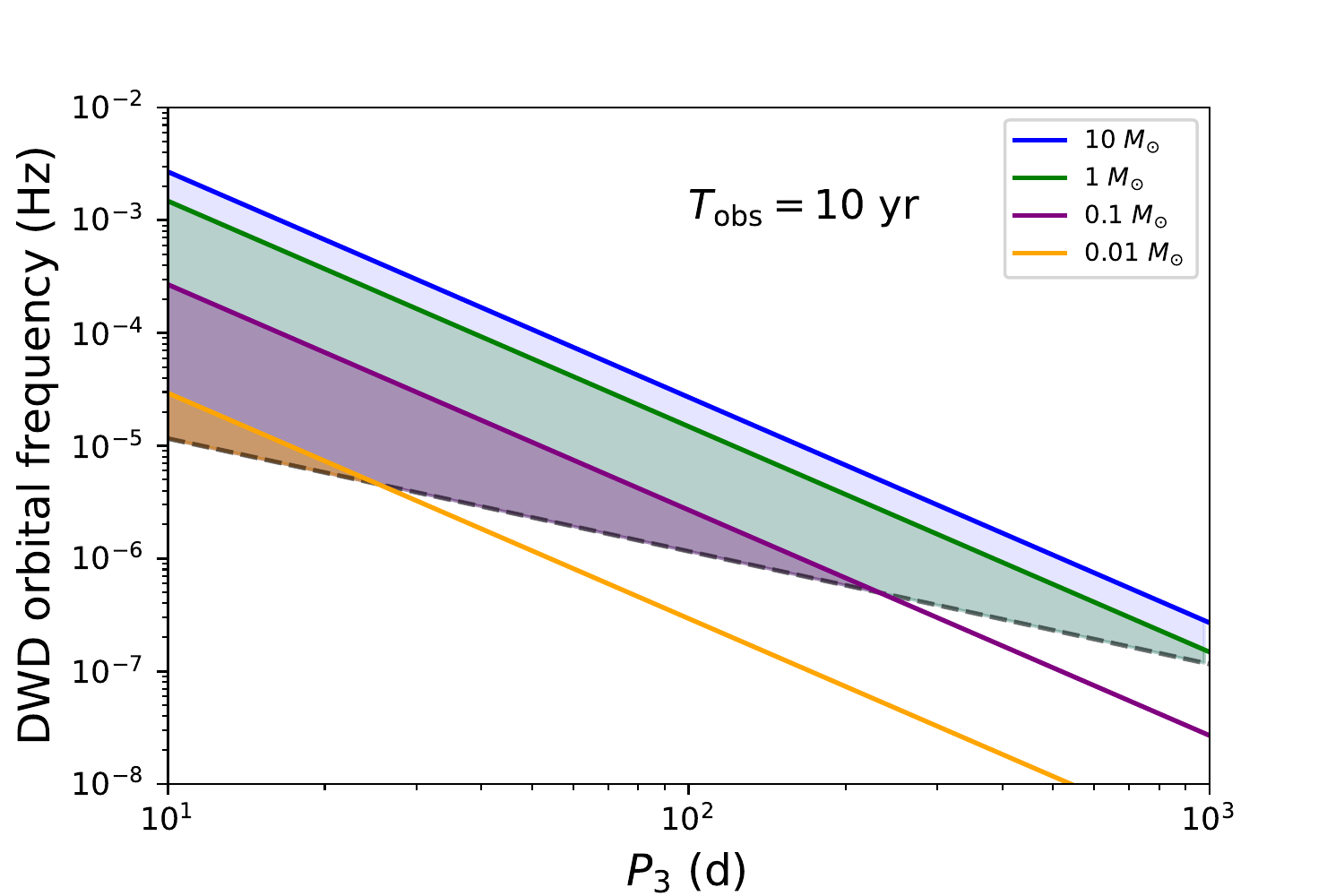}
\caption{DWD orbital frequency required to detect a third body in a four-year and ten-year LISA mission via the changing tidal field as a function of the orbital period of the third body $P_3$ and its masses of 0.01 $M_{\odot}$ (orange), 0.1 $M_{\odot}$ (purple), 1 $M_{\odot}$ (green) and 10 $M_{\odot}$ (blue). The dashed line shows the relation $P_3=10P_{12}$.  Above this line ($P_3\gtrsim10P_{12}$) the approximation of Equation \ref{eq1} is valid.  The shades indicate the region where the orbital frequency change of the DWDs are detectable by LISA. The  total mass of the DWD is 1 $m_{\odot}$ and the eccentricity of the third-body is 0.5. \label{p3f}}
\end{figure}

Assuming the total mass of the DWD is 1 $M_{\odot}$ and the eccentricity of the perturbing object is 0.5, the orbital frequency of the DWD where a third body is detectable 
is shown in Figure \ref{p3f}.  With the assumption that $P_3/P_{12}>10$, the region where the DWD frequency change is detectable by LISA is shown as the shaded region in that figure.  For a four-year observational time, a third body is detectable only around DWDs with frequencies smaller than $10^{-3}$ Hz, where we expect a low signal noise ratio of the DWD.  Moreover, detections are only possible if the orbital period of the planet is of order tens of days---where post main sequence stellar evolution may not allow stable orbits (especially stable eccentric orbits).  For planetary orbital periods of hundreds of days the necessary DWD orbital frequency drops to 10 microHertz with its accompanying drop in the GW signal-to-noise ratio.  For a ten-year observational time, the region where $m_3$ can be detected is somewhat larger.


\section{Population of the Binary White Dwarfs resolvable by LISA}\label{population}

\begin{figure}
\includegraphics[width=0.5\textwidth]{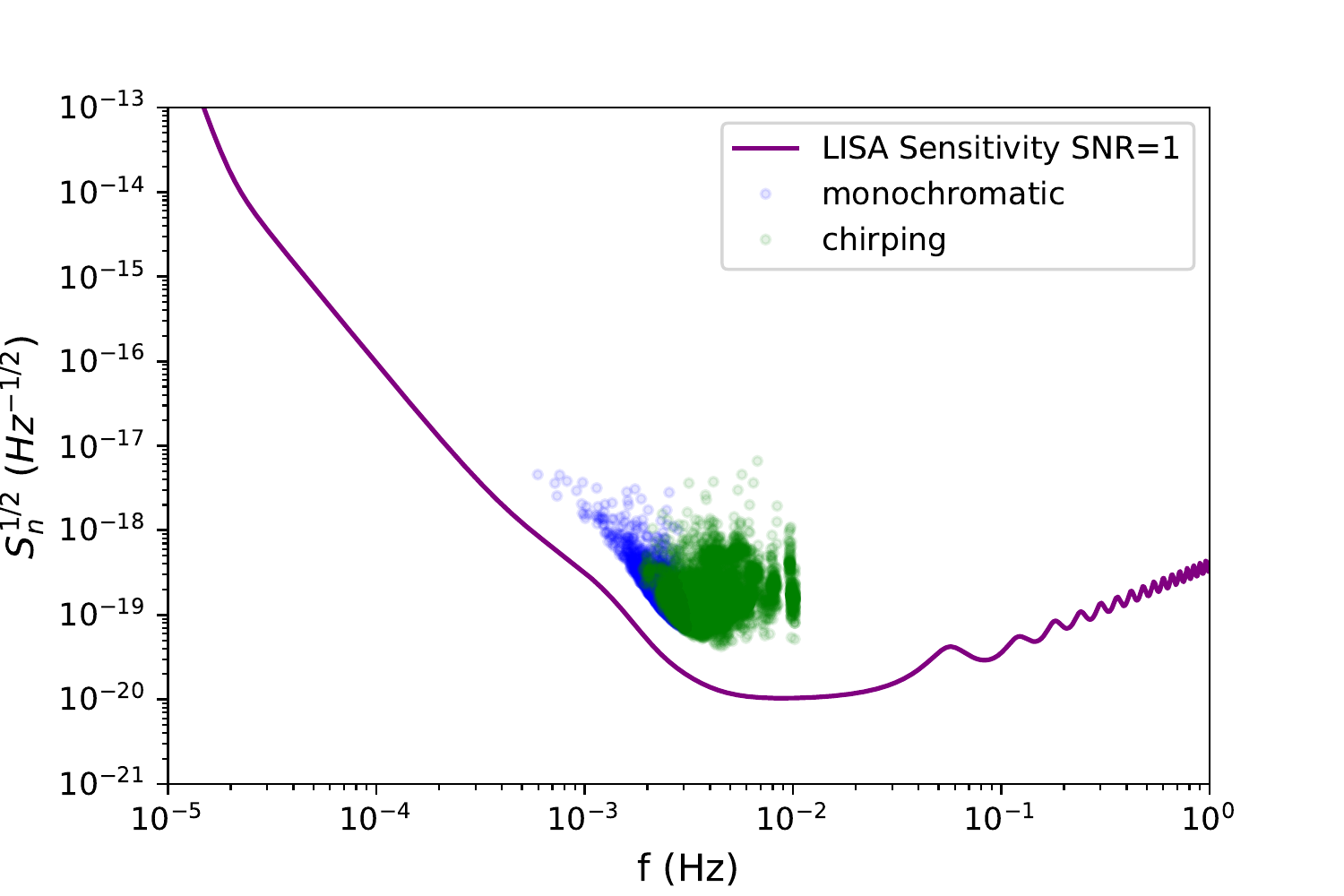}
\caption{The amplitude spectral density of the sensitivity curve combining the instrumental noise and the galactic confusion noise for a four-year mission lifetime is shown in purple. The blue dots represent resolved monochromatic DWDs, the green dots represent resolved chirping DWDs.\label{bwdsnr}}
\end{figure}

To estimate the number of circumbinary planets that might be seen in the LISA data, we begin with the population synthesis model for DWDs from \citep{COSMIC:2018}.  Assuming the strain amplitude of the gravitational wave is $h_0$, the signal-noise-ratio of the DWD can be expressed as:
\begin{equation}
S/N=\frac{h_0\sqrt{T_{\rm obs}}}{h_f}\label{snr}
\end{equation}
Where $h_f$ is the LISA sensitivity---including instrumental noise and background noise from the DWDs \citep{Cornish:2018}.  From this model, we expect about 12000 detached DWDs with a signal-to-noise-ratio $S/N>5$ over a four-year observing campaign.  This population of resolved DWDs is shown in Figure \ref{bwdsnr}.  Since the physics of DWDs with mass transfer is quite complicated, we only consider detached DWDs for our purposes.  The resolvable DWDs have GW frequencies between $10^{-3}$ Hz and $10^{-2}$ Hz.

For the tidal modulation mechanism, since the resolvable DWDs have orbital frequencies between $5\times10^{-4}$ Hz and $5\times10^{-3}$ Hz, only companions with orbital periods $<15$ days and masses $> 0.1$ $M_{\odot}$ would be detectable.  Therefore, we expect this mechanism to yield few if any detections with LISA.  We, thus, focus on the Doppler mechanism for our detection estimates for the remainder of this manuscript.

\begin{figure}
\includegraphics[width=0.5\textwidth]{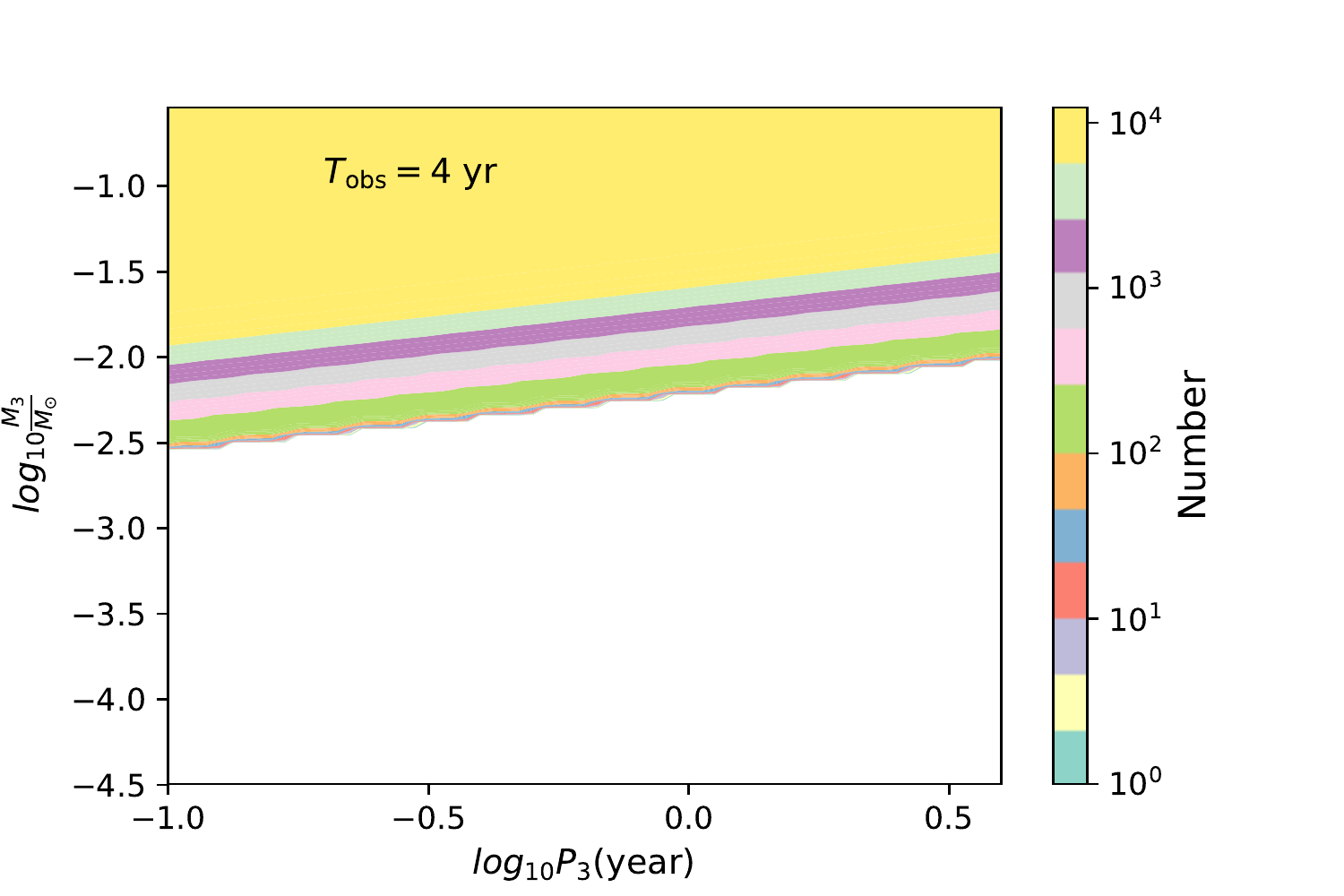}
\includegraphics[width=0.5\textwidth]{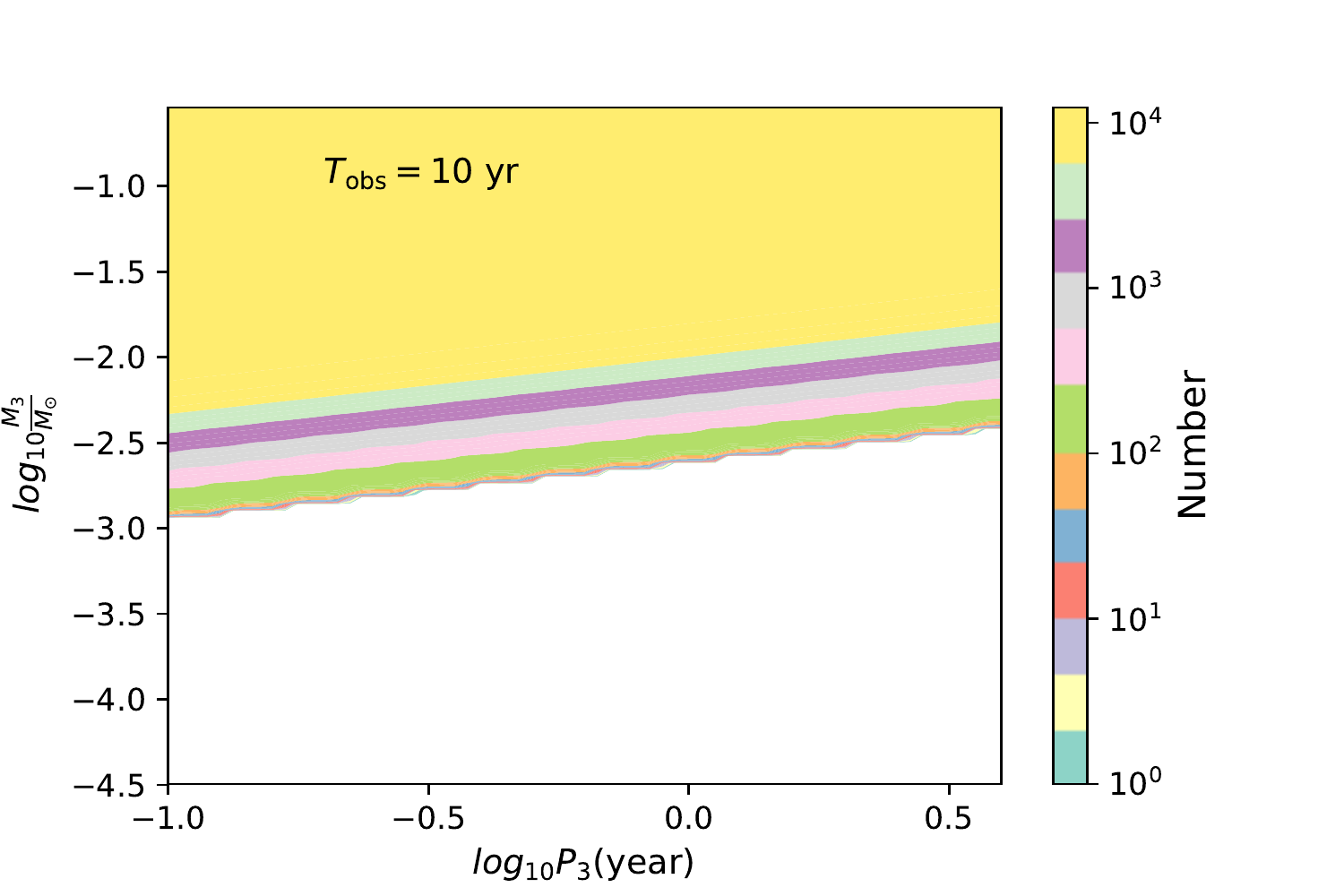}
\caption{The number of DWDs around which we can detect a third body with mass $M_3$ and orbital period $P_3$. \label{prob}}
\end{figure}

Based on the sample of resolvable DWDs, we calculate the number of DWDs $N(M_3,P_3)$ around which we can detect a companion with mass $M_3$ and orbital period $P_3$.  We assume observations spanning both four years and ten years.  The results of this calculation are shown in Figure \ref{prob}.  Giant planets or brown dwarfs with masses $M>10$ $M_{\rm J}$) and orbital periods $P_3<100$ day would be detectable around almost all resolvable DWDs.  (Note that we use the same sample of resolved DWDs for the four-year and ten-year mission since only a few outliers will have $S/N<5$ for a four-year mission and $S/N>5$ for a ten-year mission (especially at frequency $<10^{-3}$ Hz, \citet{Cornish:2017}).)

To complete our estimate, we must assume a planet occurrence rate around DWDs.  
Since the circumbinary planet occurrence rates are not well understood, we work with the assumption that their occurrence rates are similar to planets orbiting isolated solar-type F, G, and K stars.  \citet{Cumming:2008} fitted the probability density for planets with mass $>0.1$ $M_{\rm J}$ and period $<$ 2000 days around FGK stars as :
\begin{equation}
f(M,P)=CM^{\alpha}P^{\beta}d\ln M \ d\ln P
\end{equation}
Where $\alpha=-0.31\pm0.2$, $\beta=0.26\pm0.1$, $C$ is a constant.  Integrating $f(M,P)$ over a range of masses and periods yields the planet occurrence rate $F(M,P)$.  The number of the companions we can detect around the DWDs would be $N(M,P)F(M,P)$.  Since $F(M,P)$ is calculated over a range of masses and orbital periods, $N(M,P)$ is calculated as the median value of $N(M_3,P_3)$ over the same range of mass and period.  The results are shown in Table \ref{tab1}.

For a four-year observation time, we expect to see $\sim 10$ Giant planets with mass $>2 M_{\rm J}$.  Smaller giant planets will be more difficult to find.  However, when the observational time increases to ten years, we expect to find tens of giant planets with orbital periods $<100$ days and several giant planets with orbital period between $100$ days and one year.

\begin{table*}
\begin{center}
\caption{Expected number of detectable planets with a four- and ten-year observational campaign with LISA.\label{tab1}}
\resizebox{\linewidth}{!}{%
\begin{tabular}{ccccc}
\hline
&\multicolumn{2}{c|}{Four Year}&\multicolumn{2}{|c}{Ten Year}\\ 
&$2M_{\rm J}\leq M_3<10M_{\rm J}$&$10M_{\rm J}\leq M_3<15M_{\rm J}$&$2M_{\rm J}\leq M_3<10M_{\rm J}$&$10M_{\rm J}\leq M_3<15M_{\rm J}$\\\hline \hline
$P_3<100$ day&$3\pm1$&$8\pm5$&$39\pm13$&$26\pm15$\\ 
100 day $<P_3<$ one year&0&$2\pm1$&$8\pm2$&$14\pm7$\\ 
one year $<P_3<$ two year&0&0&$2\pm1$&$6\pm3$\\\hline

\end{tabular}}
\end{center}
\end{table*}

\section{Conclusions}\label{conclude}

We expect the LISA gravitational wave observatory to find several to several dozen circumbinary planets over the course of its anticipated lifetime.  These planets will be orbiting double white dwarf binaries and virtually all of them will be detected through the induced Doppler shift of the binary and its associated modulation of the GW signal.  While the modulating tidal force on the binary from an eccentric orbit of a perturbing third body does induce a GW signal, it will most often fall outside the sensitivity of the LISA mission.

These circumbinary planets will be widely distributed as LISA will be sensitive to DWD binaries everywhere within the galaxy.  Nevertheless, the detections will likely concentrate near the galactic center where there are more DWDs and the loudest examples will be somewhat closer to Earth.  For our estimates we assumed an occurrence rate similar to single stars and that the orbiting planets will survive the post Main Sequence evolution of the DWD.  This latter assumption warrants further study that lies beyond the scope of this letter.


An advantage of the approach outlined here for detecting CBPs is that the detection sensitivity of LISA is much easier to characterize than the sensitivity of transit surveys like the \kepler\ mission.  With \kepler, the constantly-changing transit profile within each system is a significant challenge to automated searches.  By contrast, the GW signal of a circumbinary planet orbiting a DWD is straightforward.  And, while the anticipated yield may be modest, it will still be useful for measuring the CBP occurrence rate across the galaxy.

\section*{Acknowledgements}

JHS is supported by NASA grant NNX16AK08G. Donghong Wu acknowledges support from the China Scholarship program.






\bibliographystyle{mnras}
\bibliography{cite}

\end{document}